\documentclass[prd,twocolumn,amsmath,amssymb,superscriptaddress,nofootinbib]{revtex4}

\usepackage{graphicx}
\usepackage{dcolumn}
\usepackage{bm}

\def\persec{{\rm s}^{-1}}

\def\eV{{\rm eV}}

\def\MeV{{\rm MeV}}
\def\GeV{{\rm GeV}}

\def\cm{{\rm cm}}
\def\sr{{\rm sr}}

\begin{document}

\title{The impacts of dark matter particle annihilation on recombination and \\
the anisotropies of the cosmic microwave background}

\author{Le Zhang}
\affiliation{National Astronomical Observatories, Chinese Academy of
  Sciences, Beijing, 100012, China}
\affiliation{Department of Physics, Shandong University, Jinan,
  250100, China}
\author{Xuelei Chen}
\email{xuelei@bao.ac.cn}
\affiliation{National Astronomical Observatories, Chinese Academy of
  Sciences, Beijing, 100012, China}

\author{Yi-An Lei}
\affiliation{Department of Physics, Peking University, Beijing, 100871, China}
\author{Zong-guo Si}
\affiliation{Department of Physics, Shandong University, Jinan,
  250100, China}

\begin{abstract}
The recombination history of the Universe provides a useful tool for 
constraining the annihilation of dark matter particles.  
Even a small fraction of dark matter particles annihilated 
during the cosmic dark age can provide sufficient energy to affect the 
ionization state of the baryonic gas. 
Although this effect is too small for neutralinos, 
lighter dark matter particle candidates, e.g. with mass of 1-100 MeV,
which was proposed recently to explain the
observed excess of positrons in the Galactic Center, may generate  
observable differences in the cosmic microwave background
(CMB) temperature and polarization anisotropies. The annihilations 
at the era of recombination affects mainly the CMB anisotropy at 
small angular scales (large $\ell$), and 
is distinctively different from the effect of early reionization. 
We perform a multi-parameter analysis of the CMB data, including
both the WMAP first year and three year data, 
and the ACBAR, Boomerang, CBI, and VSA data. Assuming that the 
observed excess of $e^+e^-$ pairs in the galactic center region 
is produced by dark matter annihilation, and that 
a sizable fraction of the 
energy produced in the annihilation is deposited in the 
baryonic gas during recombination, we obtain a \%95 
dark matter mass limit of $M<8 \MeV$ with the current data set.
\end{abstract}

\maketitle 

\section {Introduction}

The presence of dark matter, and much of what we now know 
about it, were derived from creative analysis of astronomical
observations. For example, from the abundance of the light elements, 
it has been deduced that the dark matter must be non-baryonic; and
from the large scale structure of galaxies, the 
hot dark matter candidates such as massive neutrinos were excluded.
Currently, various astronomical observations, from gravitational
lensing to the composition of cosmic rays are being studied in searches
of the dark matter \cite{BHS04}. 

The ionization history of the Universe provides us with a very useful
tool to investigate the properties of dark matter. According to the
Cold Dark Matter model with a cosmological constant ($\Lambda$CDM) model, which is
now standard in cosmology, at redshifts of about 1000, 
the temperature of the radiation background
photons was lowered sufficiently that the free electrons and protons 
could recombine to form neutral hydrogen atoms. As the number of free
electrons decreased, the gas became transparent, and most of the radiation
background photons scattered for the last time. This epoch of
recombination marks the end of the hot Big Bang and the beginning of the  
so called cosmic dark age, and much information about the last
scattering surface is preserved in the CMB anisotropies. 
The dark age lasted until the first stars formed by growth of 
primordial density
fluctuations. Eventually, the light emitted by the galaxies reionized
the Universe \cite{M03,BL01}. 

However, the history of the Universe would be different, if the dark
matter particles played a more {\it active} role during the cosmic dark age.
If the dark matter particles could decay or annihilate, 
extra energy would be injected into the baryonic gas. This 
could delay recombination, or make the Universe reionize earlier. 
These effects are observable with high precision CMB data.    
In 2003, the Wilkinson Microwave Anisotropy Probe (WMAP) 
team published the result of their first year observation
\cite{B03,H03,S03,K03,V03,Pa03}.
A strong correlation of the temperature and the E-like
polarization anisotropy (TE) was observed at large angular
scales (small $\ell$) \cite{K03,Pa03}. Such correlation could be generated by the
scattering of the CMB photons by free electrons after the
reionization of the Universe \cite{Z97,Ka03,Ho03,Pa03}. 
The best fit model requires reionization to happen at redshift 20, 
which is much earlier than predicted by the
$\Lambda$CDM model \cite{FK03,C03,Y03,So03,HH03}. 
A number of researchers has suggested that the 
decay of dark matter particle could make the reionization happen
earlier, which helps to explain the WMAP result 
\cite{CK04,BMS03,D03,HH04,P04,KKS04,PF05}. Alternatively,
using the CMB data, one could constrain the decay property of the dark
matter particle.  
As the energy corresponding to the rest mass of the dark
matter particle is much higher than the ionization energy of the
hydrogen atom, even if only a very small fraction of dark matter particles
decayed, it could inject sufficient energy to the baryonic gas to 
alter the ionization history, and affect the CMB anisotropies. 
This can be used to exclude the short-lived decaying particles with 
life time comparable to the age of the Universe at the epoch of
recombination \cite{CK04}.

In the present work, we consider the impact of dark matter
annihilation on the recombination process. Many dark matter candidate
particles could annihilate and produce $\gamma$-ray photons, energetic
electrons and positrons, and hadronic particles which ionize the
gas in the environs. For example, the annihilation processes of the
supersymmetric dark matter candidate neutralino have been well 
studied \cite{JKG96}. For neutralinos, however, the 
annihilation rate is fairly small. Since the annihilation rate is
proportional to the squared number density of the dark matter
particle, lighter particles would produce stronger annihilation
signals. Recently, a 511 keV emission line 
in the direction of the Galactic
Center was observed by the SPI spectrometer on board of 
the INTEGRAL satellite \cite{K05}. This discovery indicates the presence
of large amount of positrons in that region. It has been pointed 
out that if the dark matter
particle is not neutralino but a light scalar particle with mass 
of $1-100$ MeV and weak interaction cross sections \cite{B04,HW04},
then the annihilation rate would be high enough to produce these
positrons. Other models which attempts to explain this with dark
matter include the decay of dark matter particle \cite{KT05} or 
the annihilation of relic heavy neutrino with long
range interaction \cite{BKLS05}.

At the epoch of recombination ($z \sim 1000$), the annihilation 
rate could be even greater due to the 
higher densities, and then it might make an imprint on 
the recombination history. In Refs. \cite{PF05,MFP06} this
effect was illustrated with a few models. However, no concrete limit
on dark matter annihilation has been obtained with the current CMB
data. To obtain such a limit, one needs to calculate the CMB anisotropy 
with dark matter annihilation and compare it with the data. In doing
so, it is of crucial importance to break the degeneracies
among the many cosmological parameters, because all of these
parameters affect the CMB angular power spectrum in different ways,
and a change in the power spectrum caused by one parameter might be 
compensated by the combined variation of several other parameters.  

In the present work, we break the degeneracies among the cosmological
parameters by exploring the multi-dimensional parameter space with 
the Markov Chain Monte Carlo technique \cite{G97,KCS01,cosmomcpaper}. 
We modify the publicly available MCMC code {\tt COSMOMC} 
\cite{cosmomcpaper,cosmomcweb}, which uses {\tt CAMB}
\cite{cambpaper,cambweb} as its CMB
driver, particularly the ionization evolution code 
{\tt RECFAST} \cite{SSS99} embedded in it, to take into account
the effect of energy injection due to dark matter annihilation. We
also compare the MCMC result with the Fisher matrix estimate, to gain insights on how 
reliable is the latter method. We then use the Fisher matrix method to
make forecasts on the potential of future experiments such as the 
Planck mission \cite{Planck,Planckweb}.

In the first version of this paper, we
used the first year WMAP data\cite{B03,H03,S03,K03,V03,Pa03}, 
as well as the data obtained by the
ACBAR \cite{acb}, Boomerang \cite{Mo05}, CBI \cite{cbi}, and VSA \cite{vsa}
experiments in our analysis. Shortly after its
submission, the WMAP team released their three year observation 
data \cite{S06,Pa06,H06,J06,WMAP3}. The error of the observation is 
reduced. Also, it is now believed that the large TE correlation at low $\ell$ 
observed in the first year is due to contamination by foreground. The current
estimate of the reionization optical 
depth $\tau$ is much smaller. We have repeated our analysis with 
this new data set, and found that with the reduced errors in the three
year data, the constraint on dark matter annihilation is much stronger. The
change on $\tau$ does not significantly affect our result, because as
we shall discuss below, our constraint comes mainly from the epoch of
recombination, not the epoch of reionization.  
In this second version, we retain some results obtained in the first
version, but added new results obtained with the 
new data. 

In the following, we describe our method of calculation  
in \S II, and present our results in \S III, with a synthesis of 
current constraint on MeV dark matter obtained with 
different methods. We summarize our results in \S IV.

\section{Methods}

The evolution of the ionization fraction and temperature of the
baryonic gas is given by Eq. (15) and Eq. (23) of Ref.~\cite{CK04}, 
which we reproduce here:
\begin{eqnarray}
(1+z)\frac{dx_e}{dz}&=&\frac{1}{H(z)}[R_s(z)-I_s(z)-I_\chi(z)],\\
(1+z)\frac{dT_b}{dz}&=&\frac{8\sigma_Ta_RT^4_{cmb}}{3m_ecH(z)}
\frac{x_e}{1+f_{He}+x_e}(T_b-T_{cmb})\nonumber\\
&&-\frac{2}{3k_BH(z)}\frac{K_\chi}{1+f_{He}+x_e}+2T_b.
\end{eqnarray}
Here $R_s$ is the standard recombination rate, $I_s$ the ionization rate by
standard sources, detailed discussion of these terms can be found in 
Ref.~\cite{CK04}. The extra ionization and heating terms due to dark
matter annihilations are given by 
\begin{equation}
I_\chi = \chi_i f \frac{\Gamma_{ann}}{n_b} \frac{2 m_\chi c^2}{E_b},
 \qquad K_\chi= \chi_h f \frac{\Gamma_{ann}}{n_b} \frac{2m_\chi c^2}{E_b};
\end{equation}
where 
$n_b$ the is baryon number per unit proper volume, and
$E_b =13.6~ \eV$ is the ionization energy. 
A more detailed treatment would include helium, 
but as the spectrum of the injection energy is
quite uncertain, we will not dealt with these complications in this paper.
$\Gamma_{ann}$ is the annihilation rate. The factor $f$ 
is a fudge factor denotes the fraction of the total energy which is deposited
in the baryonic gas {\it in situ}
(c.f. Ref.~\cite{CK04}), with $f_{max}=1$.
The absorbed energy contributes both to the ionization and heating of
the gas. As a simple model for the division between these, we assume
the respective fractions are given by \cite{CK04} 
\begin{eqnarray}
\chi_i = (1-x_e)/3, \qquad \chi_h = (1+ 2 x_e)/3
\end{eqnarray}
where $x_e$ is the fraction of free electrons. 
The annihilation rate of the dark matter is given by 
\begin{equation}
\label{eq:annrate}
\Gamma_{ann}=g n^2 \langle \sigma v \rangle = g
\left(\frac{\rho_c}{m_{\chi}}\right)^2 \Omega_c^2 (1+z)^6 \langle \sigma v \rangle
\end{equation}
where $g$ is a degeneracy factor, 
$n$ is the number density of the particle, and the angular
brackets denote thermal average. The second equality applies in the
case of homogeneous distribution. 
where $\rho_c$ is the critical density at $z=0$, 
$\Omega_c$ is the relative density fraction of
the dark matter. For Majorana
particles (i.e. the particle and anti-particle are the same), 
$g=1/2$. If the particles are not Majorana, we shall assume
that the dark matter is made of equal numbers of particles and
anti-particles, and $g=1/4$. The effect of dark matter
annihilation on the ionization is then entirely quantified by a
parameter which characterizes the 
annihilation intensity: 
\begin{equation}
\label{eq:F_26}
F_{26}=2g f \left(\frac{\langle\sigma v\rangle}{10^{-26}\cm^3 \persec}\right)
(\frac{m_\chi c^2}{\GeV})^{-1}.
\end{equation}

\begin{figure}
\includegraphics[width=0.4\textwidth]{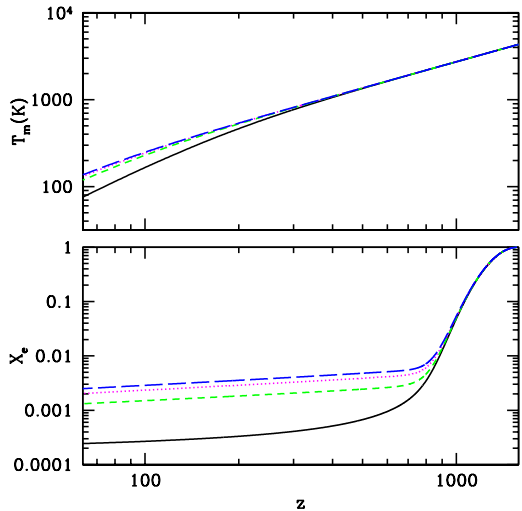}
\caption{\label{fig:F_xe_Tmat_o}The ionization fraction $x_e$ and
  intergalactic medium 
temperature
 as a function of redshift z for model (a): the fiducial model which
 is the WMAP first year best fit $\Lambda$CDM model with no contribution 
from dark matter annihilation(black solid curve); 
for (b):  $F_{26}=1.0 $ (green dotted curve);
for (c): $F_{26}=2.6$ (magenta dashed curve);
and for (d): $F_{26}=4.0$ (blue dotted - dashed curve).
We only change $F_{26}$ and keep other parameters fixed. }
\end{figure}

We modify the recombination code {\tt RECFAST} \cite{SSS99} 
to take into account of these extra contributions. For details of such
modification, see Ref.\cite{CK04}. The recombination
history for several different values of $F_{26}$ parameter 
is shown in Fig.~\ref{fig:F_xe_Tmat_o}. As can been seen from the
figure, with dark matter annihilations,
the recombination process is slightly delayed and more extended. 
This is increasingly apparent for greater value of $F_{26}$. 
However, unlike the case of 
dark matter decay investigated in Ref.~\cite{CK04} where the
ionization fraction could increase at lower redshift, the ionization
fraction still decreases steadily to an asymptotic value at later time,
because the annihilation rate drops as the number density
of the dark matter particle drops with the expansion of the
Universe. At the same time, the temperature of the gas is also
slightly higher, but still decreases steadily, instead of raising
drastically as in some decaying dark matter models. 

\begin{figure}
\includegraphics[width=0.4\textwidth]{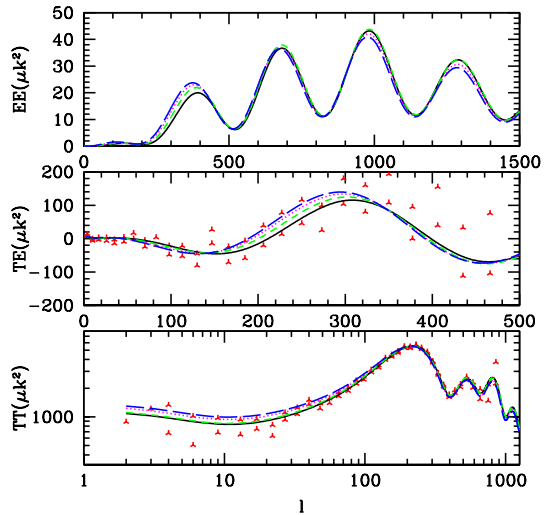}
\caption{\label{fig:p_F_o} The CMB angular power spectra for the
models (a), (b), (c) (d)given in the previous figure, plotted along with the 
CMB data points with error bars used in our fit(for WMAP: first year
data only).}
\end{figure}

The increase in free electron density 
may help to boost the formation of molecule
hydrogen, which is the most important coolant at the end of dark age.
This might increase the formation rate of the first generation of
stars. At the same time,
the increase in gas temperature also raise the Jeans mass scale, 
and thus suppress star formation in smaller dark
matter halos. The impact of dark matter annihilation on the formation
of first generation of stars is therefore interesting and
complicated, and we plan to investigate this problem in future works. 
With powerful 21cm interferometer arrays, it might be possible 
to have direct observations of gas temperature during the cosmic
dark age \cite{LZ04,BL05,BA04,FOP06}.

At present, however, the only available probe of this early epoch of
the Universe is CMB anisotropy.
The {\tt RECFAST} code was used by the Boltzmann CMB code {\tt CAMB}
\cite{cambpaper,cambweb} to
calculate the recombination history. With the above modification, we
can calculate  the angular power spectra of CMB anisotropies. The
spectra for the models described above are shown in
Fig.~\ref{fig:p_F_o}. In the TT spectrum, the amplitudes at large scale
(small $\ell$) are greater for the annihilation models. This is
because, if all parameters are kept fixed, the spectrum at large $\ell$ 
will be damped
with a factor of $e^{-2\tau}$ where $\tau$ is the optical depth
\cite{Z97,Ka03}, but when
the spectrum is fitted to the data, the greater statistical weights at
greater $\ell$ will determine the normalization of the spectrum, so the
lower $\ell$ spectrum appears to be raised \cite{CK04}. For the $TE$ spectrum,
both the position and height of the acoustic peaks are shifted, as one
might expect for a model with delayed recombination history. The
cross-correlations at small $\ell$ do not increase much in our
models, because at lower redshifts our models do not differ much
from the standard. There are also variations in the $EE$ power spectrum,
particularly in the height of the peaks.

Over the past few years, the MCMC method
has become a standard technique for exploring the multi-parameter
space, obtaining estimates on the measurement error, 
and breaking the parameter degeneracies. The publicly
available code package {\tt COSMOMC} performs such calculation with
the {\tt CAMB} code as its driver for CMB calculation. We have adopted
this package for our computation. We consider the following set of
7 cosmological parameters, ${\Omega_b h^2,\Omega_d
  h^2,\theta,\tau,n_s,A_s, F_{26}}$, 
where $\Omega_b h^2$ and $\Omega_d h^2$ are the physical density
parameters for baryon and cold dark matter particle, $\theta$ is 
the ratio of the sound horizon at recombination to its angular diameter 
distance multiplied by 100, $\tau$ is the optical depth, $n_s$ and
$A_s$ are the the spectral index and amplitude of the primordial
density perturbation power spectrum. We used the data from the 
WMAP\cite{B03,H03,K03,V03}, ACBAR \cite{acb},  Boomerang\cite{Mo05}, CBI \cite{cbi}, and VSA
\cite{vsa} experiments in our analysis. After the release of the
WMAP three year data \cite{S06,Pa06,H06,J06}, 
we repeated our analysis with the new data.

\section{Results}

\begin{figure}
\includegraphics[width=0.4\textwidth]{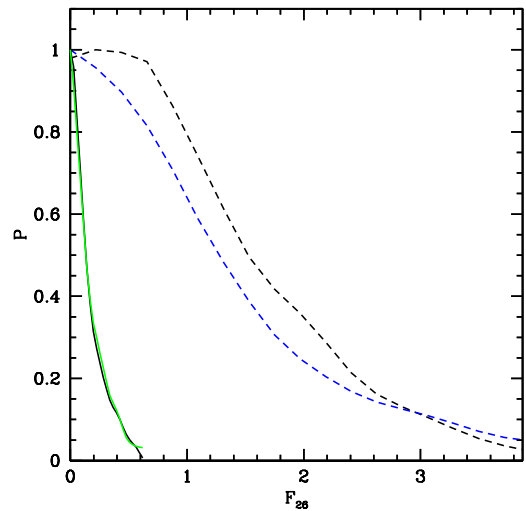}
\caption{\label{fig:1d} The marginalized probability distribution function
of the $F_{26}$ parameter and the relative mean likelihood. 
The solid curves are for WMAP three year
data, the dashed curves are for WMAP first year data. The
normalization is such that the maximum of the function is 1.
}
\end{figure}

In Fig.~\ref{fig:1d} we plot the marginalized 
probability distribution function (PDF) of the annihilation intensity
parameter $F_{26}$ and the mean relative likelihood. 
For the WMAP first year data, the PDF is fairly flat at $F \alt 1$, as 
the effect of dark matter annihilation on CMB is still too 
small compared with the measurement error at this point. 
The PDF drops more rapidly at $F_{26}> 1$, and falls below the 95\% limit
at $F_{26} \approx 2.6$.  The mean relative likelihood function (dotted
line) has a similar shape, although at $F_{26} < 1$ it falls more
steadily. This shows that our result is robust. With the WMAP three
year data, the peak of the PDF is still at $F_{26}=0$, consistent with
no detection, and the width of the PDF is much narrower, the \%95
limit is at $F_{26}=0.43$, and it dropping to 0
at $F_{26} \approx 0.6$. This indicate a significant
increase in the precision and constraining power of the new data set,
thanks in large part to the new EE power spectrum.

\begin{figure}
\includegraphics[width=0.43\textwidth]{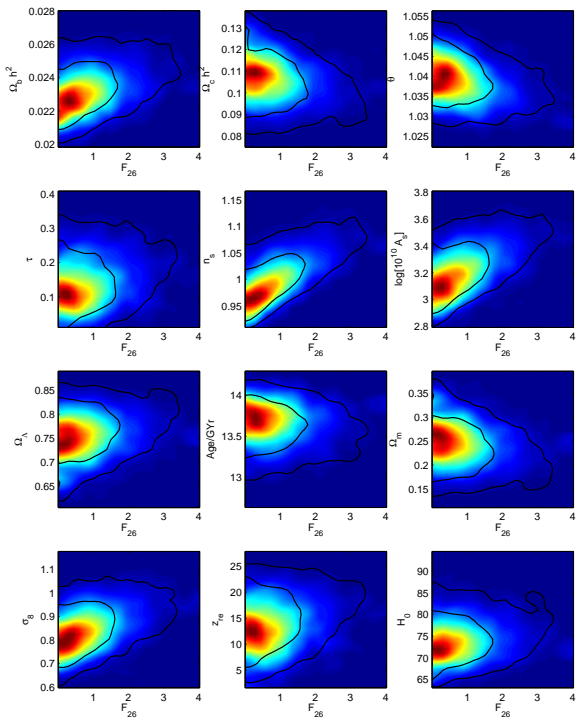}
\caption{\label{fig:2d} The $2$-D contours of the distribution of $F$ 
and background parameters for WMAP first year data. }
\end{figure}

\begin{figure}
\includegraphics[width=0.43\textwidth]{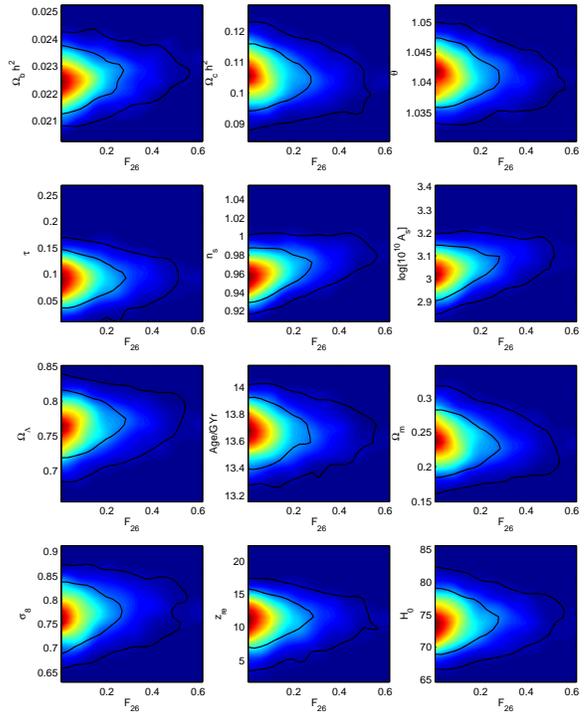}
\caption{\label{fig:2d_3} The $2$-D contours of the distribution of $F$ 
and background parameters for WMAP three year data.}
\end{figure}

How does the addition of the dark matter annihilation intensity
parameter $F_{26}$ affects the global fitting of cosmological parameters?
We list the best-fit values and errors of the cosmological parameters in
Tab.~\ref{tab:params1} (WMAP first year) and Tab.~\ref{tab:params3}
(WMAP three year). From the table, it seems that the mean value 
and errors of the other parameters are not significantly affected, as
compared to the case of the standard $\Lambda$CDM model. 
We also plot the 2-d
contours of the $F_{26}$ parameter with other cosmological parameters
in Fig.~\ref{fig:2d} (WMAP first year) and Fig.~\ref{fig:2d_3} (WMAP
three year). With the WMAP first year data, we find that $F$ correlates mainly with $n_s$, 
$\Omega_b h^2$ and $A_s$. One might naively expect a strong
correlation of $F_{26}$ 
with the reionization redshift $z_{re}$, and be surprised
that this is not so. However, in the dark matter annihilation model
described here, the impact on ionization fraction is strong at the very
high redshifts of the epoch of recombination, not at the lower
redshifts of reionization. Indeed, we find that there is little
variation in the low $\ell$ TE spectra for different values of the $F$
parameter. With the WMAP three year data, which has $EE$
power spectrum, the degeneracy is further reduced: there is very little
correlation with any parameter.

The quality of the CMB data is going to be improved continuously. How
is the $F_{26}$ parameter going to be constrained with future data,
e.g. those obtained with the Planck? To make forecasts on the measurement 
error, we use the Fisher matrix formalism (see e.g. Refs.~\cite{T97,S97,ZS97,KKS97,EHT99}).
 The Fisher matrix is computed with 
\begin{equation}
F_{ij} = \frac{\partial \chi^2}{\partial \theta_i \partial \theta_j} 
\end{equation}
and 
\begin{equation}
\chi^2 = \sum_{X} \sum_{l} 
\frac{(C_{X,l}^{obs} - C_{X,l}^{th})^2}{\sigma^2_{C_{X,l}}}
\end{equation}
where $X = TT, TE, EE, BB$. In our calculation 
we assumed a sky coverage factor of 0.65, and we adopt a fiducial
model which best fit the WMAP three year data. 
For the detector noise, we adopt the values given
in Ref.~\cite{EHT99} for the WMAP and Planck 
satellites. We have also calculated the Fisher matrix
corresponding to three years of observation of the WMAP, and found
that the result is in general agreement with that obtained with the
MCMC method, as shown in the last column of
Tab.~\ref{tab:params1},\ref{tab:params3}. There are some small residue 
differences. Given that we have also used data from several other
experiments, and the error of the WMAP is more complicated than our 
simple model with Gaussian beam, such differences are not unexpected.
We conclude that the Fisher matrix estimation is basically reliable.
The $1-\sigma$ error on
cosmological parameters calculated with the
Fisher matrix formalism for one year Planck observation is given in
Table.~\ref{tab:fisher}. 
The expected error on the $F_{26}$ parameter is 0.031.

\begin{figure}
\includegraphics[width=0.44\textwidth]{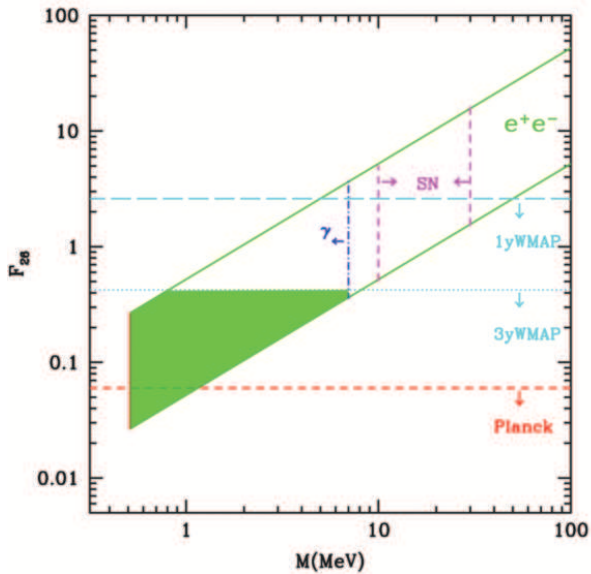}
\caption{\label{fig:constrain_all} The CMB constraint on the 
dark matter annihilation. The region between the two tilted lines
are the region required for explain the positron excess in the Galactic Center.
The lower and upper line corresponds to $f=0.1$ and $f=1$,
respectively. The
regions favored by Supernovae cooling/neutrino emission and 
continuum $\gamma$-ray flux are also marked.}
\end{figure}

The primary results of this paper is presented in
Fig.~\ref{fig:constrain_all}, which is a synthesis of constraints on 
the light dark matter annihilation flux parameter $F_{26}$ derived
from the CMB as well as other methods. 
The three horizontal lines indicate \%95 limits
on $F_{26}$ derived from CMB data. The upper two of these are derived
from the analysis of the WMAP first year and three data, together with
data from other current CMB experiments, using the MCMC method. 
The lowest one is our forecast
on the potential limit derived from Planck one year observation using
the Fisher matrix formalism. The
impact of particle annihilation on CMB depends only on $F_{26}$, so
these limits are independent of mass $M$.  

The parameter space of the light dark matter is also constrained by 
other observations. 
In particular, the original motivation of the light dark matter is to
explain the excess of positrons in the Galactic Center. 
The observed 0.511 MeV photon flux is $(1.05\pm 0.06) \times 10^{-3}
{\rm ph}~ \cm^{-2} \persec$ with an extension of $8^{\circ}$ \cite{K05}. 
The $0.511 \MeV$ photons are produced in the prompt annihilation of
positron and the $2\gamma$ annihilation of positronium. For the 
positrons in the Galactic Center, observation indicates that 
the fraction of positronium
formation is $f_p=0.96$ \cite{J05}, 
and the $2\gamma$ branching ratio of
positronium is 0.25, so for each positron the number of 0.511 MeV
photon produced is $2p$ where $p=1-0.75 f_p$. The 0.511 MeV photon
flux is then related to the dark matter annihilation cross section by 
\begin{equation}
\Phi \cong 2 p g  \bar{J} \times 5.6\, 
\bigg(\frac{\sigma v}{10^{-26} \cm^3 \persec} \bigg) 
\bigg(\frac{M}{\MeV}\bigg)^{-2}\,
\rm{cm}^{-2} \rm{s}^{-1} \sr^{-1},
\end{equation}
where $\bar{J}(0.015\sr)=231.8$ \cite{A06} 
for the Navarro-Frenk-White profile \cite{NFW97}.
Thus, to produce the observed positrons by dark matter annihilation,
the cross section of the dark matter is given by
\begin{equation}
\frac{\langle \sigma v \rangle}{10^{-26} \cm^3 \persec} = 7.8 \times 10^{-5} 
g^{-1} p^{-1} \left(\frac{M}{\MeV}\right)^2.
\end{equation}
In terms of $F_{26}$, 
\begin{equation}
F_{26} = 0.156 p^{-1} f \left(\frac{M}{\MeV}\right).
\end{equation}
Assuming $0.1<f<1$, we draw the favored region as the tilted band
marked $e^+e^-$ (raising from left to right). The band is bounded from
left, as the mass of the dark matter particle must be greater than
0.511 MeV to produce electron-positron pair in its annihilation.
By combining this requirement with the CMB bound on $F_{26}$, we can
derive an upper bound on the dark matter particle mass. 
The uncertainty (width) of this $e^+e^-$ band is mainly due to the
uncertainty in $f$, hence the 
mass upper bound also depends on $f$. 
If we assume value $f \simeq 1$, i.e. a
large part of the energy released during annihilation could 
contribute to the ionization process, which 
corresponds to the upper border of the $e^+e^-$ band. 
In this case, even with the first year WMAP data, an upper bound of 
5 MeV on the dark matter can be obtained. 
On the other hand, for small $f \sim 0.1$, the upper bound on the mass
obtained with the WMAP first year data is 
about 50 MeV \footnote{
  These constraints are stronger than given in the first draft of this
  paper. In the first draft we
  assumed that all positrons annihilate in the two photon
  process. However, a large fraction of positrons would form
  positronium, and annihilate in three photon process, hence to
  produce the observed flux, greater annihilation rate is required. 
  Also, in the first draft we made a mistake in
  converting units used in some literature to 
  the units used in this paper.}. 
With the WMAP three year data, the allowed region is further 
reduced. For $f=1$, the upper bound on mass is only 0.8 MeV, which
excludes most of the allowed mass range.
Even for $f=0.1$, we still obtain a strong limit of $M < 8 \MeV$. As we assume
that the $e^-e^+$ pair is produced by dark matter annihilation
(i.e. not invisible decay), it is unlikely for the value of $f$ to be
much smaller.  This limit can be further improved with future high precision
experiments such as the Planck Surveyor. The \%95 limit for one year 
observation of Planck(assuming the same set of cosmological parameters) is 
$F_{26} \sim 0.06$.  For $f=1$, the whole mass range can be excluded. 
Even for $f=0.1$, this will produce an upper mass limit of about 1
MeV. The CMB limit can be evaded if one adopts a very small value of
$f$. This would happen if, e.g., weakly interacting annihilation
products such as neutrinos carry most of the energy away. Such
annihilating dark matter would have little impact on baryonic gas, but
could be constrained with neutrino detectors \cite{BBM06}.

Beside the CMB and positron excess, the light dark matter can also be
constrained with $\gamma$-ray emission from 
annihilation \cite{BY05,BBB05,SCS06,KE05}, and from the cooling and neutrino
emission of the core-collapse supernovae explosion \cite{FHS06}. 
The mass range of 10-30 MeV is favored by the SN argument,
although this value can be lowered if the dark matter-neutrino
interaction is not enhanced as expected. The limit derived from the 
$\gamma$-ray flux depends somewhat on the model and data adopted,
ranging from 3 MeV \cite{BY05} to 20 MeV \cite{KE05}.  
On Fig.~\ref{fig:constrain_all}, we marked $M < 7.5 ~\MeV$ as derived
from Ref.~\cite{SCS06} as the $\gamma$ limit.
The favored regions derived with these
methods do not overlap. The higher mass range favored by the supernova
argument is also difficult to reconcile with the relic abundance of
the dark matter, if it is produced by the conventional thermal
mechanism. Our result based on the WMAP three year data is 
compatible with the $\gamma$-ray limits.

\section{Conclusion}

We have investigated the effect of dark matter annihilation on
recombination history and the CMB anisotropies. Because the
annihilation rate is only significant at the high redshifts of the era
of recombination, the shift in the CMB angular power spectra occurs 
mostly at high $\ell$s, and the signature is distinct from that of an
early reionization. The impact of the energy injection from
annihilating dark matter on the ionization 
history and CMB spectra can be characterized by the 
annihilation intensity parameter, $F_{26}$ (c.f. Eq.~\ref{eq:F_26}). 
The addition of this parameter does not significantly affect the 
uncertainty in the estimation of the other parameters. 
Using the currently 
available data, including WMAP first year and three data, as well as
those from ACBAR, Boomerang, CBI and VAS, we obtained limits on this 
parameter. With the WMAP first year data, we can start to constrain the 
parameter space. Dark matter annihilation with $F_{26} > 2.6 $ is
excluded at \%95 level, corresponding to mass greater than 60 MeV. The WMAP 
three year data, particularly the EE spectrum, provides much stronger
constraint: $F_{26} < 0.43$ at \%95 limit, corresponding to $M < 8
~\MeV$. This result is compatible with the limit derived from
$\gamma$-ray observation \cite{BY05,BBB05,SCS06,KE05}, and apparently
excludes the region favored by the supernovae cooling argument \cite{FHS06}.
With future CMB data such as those from Planck,
these limits can be future strengthened, or even more
interestingly, possible signal of dark matter annihilation could be
discovered. In addition, heating of the baryonic gas to high
temperature during the dark age might be observable with future 21cm
observations \cite{FOP06}. 
Discovery of such signal would provide strong evidence 
for energy injection from dark matter decay or annihilations.

\acknowledgments

We thank Bo Feng, Xiaojun Bi, Junqing Xia, Gongbo Zhao, Quan Guo, Yan
Qu, Pengjie Zhang, Hongsheng Zhao and John F. Beacom 
for discussions and suggestions. Our MCMC chain 
computation was performed on the Shenteng system of the 
Supercomputing Center of the Chinese Academy
of Sciences. This work is supported by
the National Science Foundation of China under the Distinguished Young
Scholar Grant 10525314, the Key Project Grant 10533010, and Grant 10575004.

\begin{table*}
\caption{\label{tab:params1}
First year WMAP posterior constrains  on cosmological parameters}

\begin{tabular}{|c|c|c|c|c|c|c|c|}   \hline\hline
 Parameter &best-fit & mean	& $1\sigma$ lower & $1\sigma$ upper
 &$2\sigma$ lower &  $2\sigma$ upper    & $\sigma$(fisher) \\
\hline
 
  $100\Omega_b h^2$     &$2.226$   &$2.411$      &$2.244$ 	&$2.586 $  &$2.129$ 	&$2.872$   &$0.072$ \\

  $\Omega_c h^2$   &$0.109$    &$0.105$       &$0.0944$  &$0.114$  &$0.0850$   &$0.127$ &$0.0055$ \\

 $\Theta_S$          &$1.041$      &$1.039 $     &$1.034$    &$ 1.045$     &$1.029$      &$1.050$  &$$          \\

  $\tau$        &$0.107$      &$ 0.145$     &$0.01$    &$ 0.173$     &$0.01$      &$0.281$    &$0.0185$                  \\

  $F_{26}$            &$0.267 $       &$0.94$     &$0$    &$1.13$     &$0$      &$2.62$ &$0.317$ \\

  $n_s$        &$0.97 $       &$1.006$    &$0.962$     &$1.050 $    &$0.937$ 	&$1.106$ &$0.026$                \\

  $\log[10^{10} A_s]$  &$3.116 $    &$ 3.244$    &$3.077$     &$3.422 $    	&$2.949$  &$3.610$    &$0.0796$                    \\
  \hline\
  $\Omega_\Lambda$  &$0.745$     &$ 0.766$    &$0.723$     &$0.810 $    	&$0.663$  &$0.845$   &$$                  \\

  Age/Gyr    &$13.67 $     &$13.61 $    &$13.39$     &$13.86 $    &$13.05$ &$14.04$      &$$                  \\

  $\Omega_m$   &$0.255 $        &$0.234 $    &$0.190$     &$ 0.277$    &$0.155$ &$0.337$   &$$                      \\
  $\sigma_8$      &$0.82 $       &$0.855 $    &$0.766$     &$0.945 $    &$0.700$ &$1.046$      &$$               \\   

  $z_{re}$      &$12.68 $       &$14.84 $    &$10.00$     &$19.82 $    &$5.92$ &$23.6$     &$$                \\

  $H_0$        &$72.1 $           &$74.57 $    &$70.26$     &$78.99 $    &$66.56$ &$84.95$  &$0.0283$  \\
   
\hline \hline

\end{tabular}
\end{table*}

\begin{table*}
\caption{\label{tab:params3}
Three year WMAP data posterior constrains  on cosmological parameters}

\begin{tabular}{|c|c|c|c|c|c|c|c|}   \hline\hline
 Parameter &best-fit & mean	& $1\sigma$ lower & $1\sigma$ upper &
 $2\sigma$ lower & $2\sigma$ upper  & $\sigma$(fisher) \\              
\hline
 
  $100\Omega_b h^2$     &$2.249$   &$2.257$      &$2.192$ 	&$2.325 $ &$2.128$ 	&$2.394$  &$0.054$ \\

  $\Omega_c h^2$   &$0.1049$    &$0.1047$       &$0.099$  &$0.111$  &$0.092$   &$0.118$ &$0.0038$\\
 
$\Theta_S$          &$1.0411$      &$1.0412 $     &$1.038$    &$ 1.044$     &$1.035$      &$1.048$  &$$     \\
  $\tau$        &$0.096$      &$ 0.091$     &$0.01$    &$ 0.105$     &$0.01$      &$0.138$   &$0.0116$ \\
  $F_{26}$            &$0.007 $       &$0.147$     &$0$    &$0.174$     &$0$      &$0.424$ &$0.173$      \\
  $n_s$        &$0.956 $       &$0.964$    &$0.947$     &$0.981 $    &$0.932$ 	&$0.999$  &$0.0174$     \\
  $\log[10^{10} A_s]$  &$3.033 $    &$ 3.049$    &$2.983$     &$3.113 $    	&$2.917$ &$3.178$   &$0.043$  \\                 
  \hline\
  $\Omega_\Lambda$  &$0.767$     &$ 0.766$    &$0.738$     &$0.794 $    	&$0.704$  &$0.820$  &$$  \\
  Age/Gyr    &$13.654 $     &$13.651 $    &$13.50$     &$13.8 $    &$13.34$ &$13.92$    &$$             \\

  $\Omega_m$   &$0.2331 $        &$0.2336 $    &$0.206$     &$ 0.261$    &$0.179$ &$0.295$   &$$         \\
  $\sigma_8$      &$0.764 $       &$0.771 $    &$0.729$     &$0.812 $    &$0.689$ &$0.851$  &$$         \\   
  $z_{re}$      &$11.83 $       &$11.15 $    &$0.876$     &$13.56 $    &$5.41$ &$15.52$  &$$             \\
  $H_0$        &$73.95 $           &$74.11 $    &$71.28$     &$76.91 $    &$68.62$ &$80.40$  &$0.0227$   \\
\hline

\hline \hline

\end{tabular}
\end{table*}

\begin{table*}
\caption{\label{tab:fisher} Fisher matrix forecast on the 
1-$\sigma$ error of cosmological
parameters measured with the Planck satellite.}
\begin{tabular}{|c|c|c|c|c|c|c|c|}
\hline\hline
parameter & $\Omega_b h^2$ &$\Omega_c h^2$ & $h$ &
$\tau$ &$A_s$ & $n_s$ &$F_{26}$ \\
\hline
error ($1\sigma$)&$1.7 \times 10^{-4}$ &  $1.5 \times 10^{-3}$
 & $7.8\times 10^{-3}$ &  0.0050 &  0.012 &   0.0042 &0.031\\
\hline
\hline
\end{tabular}

\end{table*}

\newpage

\newcommand\AAP[3]{Astron. Astrophys.{\bf ~#1}, #2~ (#3)}
\newcommand\AL[3]{Astron. Lett.{\bf ~#1}, #2~ (#3)}
\newcommand\AP[3]{Astropart. Phys.{\bf ~#1}, #2~ (#3)}
\newcommand\AJ[3]{Astron. J.{\bf ~#1}, #2~(#3)}
\newcommand\APJ[3]{Astrophys. J.{\bf ~#1}, #2~ (#3)}
\newcommand\APJL[3]{Astrophys. J. Lett. {\bf ~#1}, L#2~(#3)}
\newcommand\APJS[3]{Astrophys. J. Suppl. Ser.{\bf ~#1}, #2~(#3)}
\newcommand\MNRAS[3]{Mon. Not. R. Astron. Soc.{\bf ~#1}, #2~(#3)}
\newcommand\MNRASL[3]{Mon. Not. R. Astron. Soc.{\bf ~#1}, L#2~(#3)}
\newcommand\NPB[3]{Nucl. Phys. B{\bf ~#1}, #2~(#3)}
\newcommand\PLB[3]{Phys. Lett. B{\bf ~#1}, #2~(#3)}
\newcommand\PRL[3]{Phys. Rev. Lett.{\bf ~#1}, #2~(#3)}
\newcommand\PR[3]{Phys. Rep.{\bf ~#1}, #2~(#3)}
\newcommand\PRD[3]{Phys. Rev. D{\bf ~#1}, #2~(#3)}
\newcommand\SJNP[3]{Sov. J. Nucl. Phys.{\bf ~#1}, #2~(#3)}
\newcommand\ZPC[3]{Z. Phys. C{\bf ~#1}, #2~(#3)}

\end{document}